# Accounting for electronic polarization in non-polarizable force fields

Igor Leontyev and Alexei Stuchebrukhov

*Department of Chemistry, University of California Davis, One Shields Avenue, Davis, California 95616*

**Abstract:** The issues of electronic polarizability in molecular dynamics simulations are discussed. We argue that the charges of ionized groups in proteins, and charges of ions in conventional non-polarizable force fields such as CHARMM, AMBER, GROMOS, etc should be scaled by a factor about 0.7. Our model explains why a neglect of electronic solvation energy, which typically amounts to about a half of total solvation energy, in non-polarizable simulations with un-scaled charges can produce a correct result; however, the correct solvation energy of ions does not guarantee the correctness of ion-ion pair interactions in many non-polarizable simulations. The inclusion of electronic screening for charged moieties is shown to result in significant changes in protein dynamics and can give rise to new qualitative results compared with the traditional non-polarizable force field simulations. The model also explains the striking difference between the value of water dipole $\mu$~3D reported in recent *ab initio* and experimental studies with the value $\mu^{eff}$~2.3D typically used in the empirical potentials, such as TIP3P or SPC/E. It is shown that the effective dipole of water can be understood as a scaled value $\mu^{eff} = \mu/\sqrt{\varepsilon_{el}}$, where $\varepsilon_{el}$ =1.78 is the electronic (high-frequency) dielectric constant of water. This simple theoretical framework provides important insights into the nature of the effective parameters, which is crucial when the computational models of liquid water are used for simulations in different environments, such as proteins, or for interaction with solutes.

## 1. Introduction

At present, the majority of molecular dynamics simulations are performed by using non-polarizable force fields such as AMBER[1], CHARMM[2], GROMOS[3] and OPLS[4]. In these models, the all-important effects of electronic polarization and screening[*] of electrostatic interactions are presumably incorporated in the effective charges and other empirical parameters of the force fields.

Despite the drastic simplifications, non-polarizable models have been remarkably successful in modeling many complex molecular systems.[6] For example, the properties of liquid water are described quite accurately without introducing electronic polarizability explicitly; likewise, the hydration free energies can be computed quite accurately using non-polarizable simulations.[7-8] However, the simulation of polarization effects in low-polar solvents, e.g. ethers,[9] and especially in non-polar solvents, e.g. alkanes,[10-11] meet serious problems. The non-polarizable models can also significantly underestimate the magnitude of the dielectric response in low-dielectric protein environment[12-13] and lipid membranes[14]. For example, the dielectric constant of the inner part of cytochrome *c* was found to be only about 1.5,[15] which is lower than pure electronic dielectric constant $\varepsilon_{el} \cong 2.0$.[16] Many other shortcomings of non-polarizable MD simulations have been recently discussed in the literature, see [17] and references therein.

---

[*] Throughout the paper the term "electronic screening" means a reduction of the electric field and electrostatic interactions due to an electronic relaxation of the environment. For the origin of the effect see e. g. ref.[5]

The polarizable models aim at resolving the problems mentioned above. Most of such models involve various kinds of coupled polarizable sites [9-10, 18-19, 20-22], and the computationally-expensive procedure of achieving self-consistency of such sites at each molecular dynamics time step[†]. The implementation of such models is yet to be completed; at present, even the simplest classical Drude oscillator model [9-10, 20-22] is still not readily available for application to many biological systems.

As fully polarizable force fields are being developed, there is also a clear need for better understanding the existing non-polarizable models, in particular how accurately they capture the effects of electronic polarization and screening[23-24], and possibly improving them. Given a specially designed (but empirical in nature) procedure of how the partial charges are selected [1, 2] the charges of neutral residues do reflect, at least approximately, the effects of electronic screening - in a way how for example TIP3P or similar fixed-charge models of water does so. One issue of concern, however, is that the electrostatic interactions of ions are described in standard non-polarizable force fields, such as CHARMM or AMBER, by their original integer charges (e.g. $\pm 1$, for $Na^+$ and $Cl^-$), i.e. as if these ions were in vacuum, completely disregarding the effect of electronic dielectric ($\varepsilon = \varepsilon_{el}$) screening inherent to the condensed phase medium. The interaction of such bare charges obviously is overestimated by a factor of about two (the screening factor $\varepsilon_{el}$ is about 2 for most of organic media[10]). Thus, for example, in simulation of ion channels with conventional non-polarizable force fields, the direct Coulomb interaction of ions (e.g. several $K^+$ ions in the same channel, just a few angstroms apart[25]) is probably twice as strong as it should be; similarly, the interactions of ions with water molecules, or with partial atomic charges of a protein are likely to be overestimated as well. The same is true for interaction of *charged* residues in proteins, such as $Arg^+$ or $Glu^-$, partial charges of which carry their original net values $\pm 1$. The use of bare charges in non-polarizable simulations would be appropriate for vacuum, but not for condensed phase, where all charges are essentially immersed in the electronic continuum, which weakens their interactions by a factor of about two.

---

[†] With the Extended-Lagrangian technique[9-10, 18, 20-22] the computation cost of polarizable simulations can be significantly reduced.

A similar problem arises in QM/MM calculations, where one needs to evaluate the electric field of the protein medium to which the QM system is exposed. The use of CHARMM or AMBER charges in such calculations has become standard, and has been adopted in many studies [26]. Obviously, the electric potential of charged residues in such calculations should reflect the electronic screening of the medium.

In this paper we discuss a principle of uniform charge-scaling based on which one could systematically build a non-polarizable force field for simulations of condensed media. The principle is based on a simple idea of uniform electronic continuum, with an effective dielectric constant $\varepsilon \sim 2$, and point charges moving in it. The resulting model, which combines a non-polarizable (fixed-charge) force field for nuclear dynamics (MD) with a phenomenological electronic continuum (EC) is referred to as MDEC (Molecular Dynamics in Electronic Continuum). In some sense, the model is an opposite limit of fully polarizable models involving polarizable point dipoles. Of course, the reality neither involves point dipoles nor the completely uniform electronic continuum, but lies somewhere in between these two limits. In MDEC model the effects of electronic screening are reduced to simple scaling of partial charges. The model is similar but not equivalent to standard non-polarizable force fields[1, 2-4]. An ultimate rigorous implementation of the new concept, of course, would require a consistent re-parameterization of all force field parameters such as bond-length, angle, torsion and van-der-Waals parameters along with the effective partial charges. In this review, however, we examine only a simple scaling of partial charges, which makes non-polarizable force fields such as AMBER and CHARMM to be uniformly consistent with the idea of electronic screening that naturally should improve the quality of these force fields.

Of our particular interest is the calculation of solvation effects using MD. In the MDEC the electronic polarization part of the solvation is calculated explicitly from the electronic continuum model, while the nuclear part is obtained with a fixed-charge MD. The two parts need to be combined to obtain the total solvation energy[27-28]. We will demonstrate that MDEC model and the Drude oscillator model produce comparable results for dielectric constants of alcohols and alkanes. It will be then argued that using this model one can rather accurately describe the solvation effects and dielectric properties of non-polar liquids and proteins.

Several examples of MDEC calculations and the effects of electronic polarization will be discussed, including the interaction between $Na^+$ ions, which is of interest for ion-channel simulations, and the dynamics of an important salt-bride in Cytochrome *c* Oxidase.

Another important issue related to non-polarizable models concerns water. We show that such models as TIP3P and SPC/E in effect are MDEC models. Our theory explains the striking difference between the value of bulk water dipole $\mu_l \sim 3D$ reported in recent *ab initio* and experimental studies and the value $\mu^{eff}$~2.3D of TIP3P or SPC/E. We show that the effective dipole of water can be understood as a scaled value $\mu^{eff} = \mu_l/\sqrt{\varepsilon_{el}}$, where $\varepsilon_{el}$ =1.78 is the electronic (high-frequency) dielectric constant of water. This simple theoretical framework provides important insights into the nature of the effective parameters, which is crucial when the computational models of liquid water are used for simulations in different environments, such as proteins, or for interaction with solutes.

## 2. Theory. MDEC model.

The MDEC model was discussed previously in refs [27-29]. Here we summarize the main features of the model essential for the subsequent discussion.

The MDEC model considers point charges moving in homogeneous electronic continuum of known dielectric constant $\varepsilon_{el}$. The interactions between charges in such a system are scaled by a factor $1/\varepsilon_{el}$. It is instructive to see how this model arises from a microscopic polarizable model as an approximation. A formal analysis of this is given next.

**2.1 Screening Effect and Effective Charges**

Consider a system of polarizable point charges. The energy of such a system is written as follows:

$$W(r_1,...,r_N) = \frac{1}{2}\sum_{j\neq i}^{N}\frac{q_i q_j}{r_{ij}} + \frac{1}{2}\sum_{i,j=1}^{N}d_i K(i,j)d_j - \sum_{i=1}^{N}E`(r_i)d_i \qquad (2.1)$$

where $q$'s are the gas-phase partial atomic charges, and $d$'s are the induced point dipoles located at the positions $r$'s of the corresponding charges. The dipoles are induced by the electric field from other charges and other dipoles. The dipole-dipole interaction is quadratic and is described by the matrix $K$; the diagonal elements of this matrix are inverse polarizabilities, $1/\alpha$, which are assumed to be the same for all charges. In addition to the dipole-dipole interactions, the dipoles also interact with the electric field of other point charges, $E`(r_i)$. The field is taken at the position of a given polarizable site (and corresponding charge) $r_i$, and the prime indicates that the electric field does not include the field of the point charge itself. For simplicity, hereafter, the vector notations are omitted while the usual vector nature of the appropriate variables is assumed; thus, e.g. $E`(r_i)d_i$ denotes a scalar (dot) vector product, and $d_i K(i,j) d_j$ stands for the vector tensor vector product.

The polarizable dipoles represent electronic polarizability of the ions, and therefore respond to an external field "instantaneously". The external field here is the field of point atomic charges, which is changing together with the position of the nuclei on a much slower time-scale than the electronic response. Thus the polarization dipoles are always at "equilibrium" (i.e. minimizing the total energy) for a given configuration of the nuclei, and the dynamics of the nuclei coordinates $r$ can be described with a Born-Oppenheimer type of effective potential energy $W(r_1,...,r_N)$; the dynamic coordinates of the dipoles are not present explicitly in this picture.

The equilibrium values of the dipoles can be found by minimizing the energy with respect to the dipole values. Each of the dipoles will have the following equilibrium value:

$$\bar{d}_i = \alpha\left( E'(r_i) - \sum_{j \neq i}^{N} K(i,j)\bar{d}_j \right) \qquad (2.2)$$

where the first term in parenthesis is the electric field of the charges other than $q_i$, and the second is the electric field of other dipoles $d_j$ at the position of the dipole $d_i$. All equilibrium values of the dipoles depend self-consistently on each other, and on the position of the nuclei, which determine the "external" field to which the dipoles are subjected to. The substitution of the above equilibrium values for dipoles into energy expression gives

$$W(r_1,...,r_N) = \frac{1}{2}\sum_{j \neq i}^{N} \frac{q_i q_j}{r_{ij}} - \frac{1}{2}\sum_{i=1}^{N} E`(r_i)\bar{d}_i \qquad (2.3)$$

The second term of the above equation is the energy of the dipoles, which is the same as electronic polarization energy. The point dipole polarization is now written in terms of the polarized continuum as follows:

$$\bar{d}_i = \int_{V_i} P`(r)dr \qquad (2.4)$$

where $P`(r)$ is polarization density, and the integration is over the volume $V_i$ of the $i$-th atom. Since the boundaries between atoms are not well defined, here already the approximate character of the treatment becomes evident. The prime of the polarization density indicates that this is part of total polarization at point $r$ caused by the electric field other that the field of the atom itself. This polarization is proportional to the local external field, as in the usual macroscopic continuum electrostatics (here the electric displacement $D$ is the same as $E`$):

$$P`(r) = \frac{1}{4\pi}\frac{\varepsilon_{el}-1}{\varepsilon_{el}}E`(r) \qquad (2.5)$$

Assuming now that the external electric field $E`(r)$ does not change significantly within the atomic dimensions (this is the second major approximation), the polarization energy can be written as follows

$$W_{el} = -\frac{1}{2}\sum_{i=1}^{N} E`(r_i)\overline{d}_i = -\left(\frac{\varepsilon_{el}-1}{\varepsilon_{el}}\right)\sum_{i=1}^{N} \int_{Vi} \frac{E`^2(r)}{8\pi}dr ,\qquad(2.6)$$

which after some additional transformations and assuming spherical shape of the atoms becomes

$$W_{el} = -\left(1-\frac{1}{\varepsilon_{el}}\right)\left(\frac{1}{2}\sum_{j\neq i}^{N}\frac{q_i q_j}{r_{ij}} + \frac{1}{2}\sum_{i=1}^{N}\frac{q_i^2}{R_i}\right).\qquad(2.7)$$

where $R$'s are the corresponding radii of the atoms. The substitution of the above relations into Eq.(2.3) gives for total energy the following:

$$W(r_1,...,r_N) = \frac{1}{2}\sum_{j\neq i}^{N}\frac{q_i q_j}{\varepsilon_{el} r_{ij}} - \sum_{i=1}^{N}(1-1/\varepsilon_{el})\frac{q_i^2}{2R_i}.\qquad(2.8)$$

As expected, the first term, which represents interatomic interactions, is scaled by a factor $1/\varepsilon_{el}$. The second term in this expression is the polarization (Born) energy of individual ions, which does not depend on the atomic configuration and, therefore, is not important for interaction and dynamics of the charges; however it becomes critically important in solvation energy calculations.

The above expression could have been written from the start for a system of charges in electronic continuum. The above derivation shows that the system of polarizable dipoles can be approximated by an effective non-polarizable model. It follows from the above derivation that the dielectric constant $\varepsilon_{el}$ is related to polarizability of the dipoles $\alpha$ and density of the polarizable sites $N_\alpha$ as given by the Clausius-Massotti expression:

$$\varepsilon_{el} - 1 = \frac{4\pi\alpha N_\alpha}{1-(4\pi/3)\alpha N_\alpha} \qquad (2.9)$$

Thus, the energy of a system of point polarizable dipoles can be approximated by that of an equivalent continuum model. The numerical quality of the approximation is difficult to evaluate a priori, however, despite the known steps of the derivation that involve approximations. We will check the quality of this approximation by comparing directly the results of calculations using Drude model and an equivalent continuum model later in the paper.

From the above treatment it follows that the system of polarizable point charges can be substituted by a system of non-polarizable point charges of scaled values $q_i^{eff} = q_i/\sqrt{\varepsilon_{el}}$, so that the interaction between the scaled charges correctly reproduces the actual interaction $q_i^{eff} q_j^{eff}/r_{ij} = q_i q_j/\varepsilon_{el} r_{ij}$ as if they were in vacuum.

The model described above is strictly valid only for ionic mono-atomic liquids[30]. However, if one deals with groups of charges, which represent molecules, e.g. water, instead of individual point charges, the results are formally the same as above; only in this case there is no simple relation between partial charges of molecules in vacuum and effective charges in the condensed phase. A good example of such situation is water molecule discussed in the text. It should be noticed too that in a different type of a model, in which an artificial boundary between the molecule and the rest of the electronic polarizable continuum is introduced, for the case of neutral molecules the screening factor may differ from $1/\varepsilon$ and depend on a shape of the molecular cavity[‡], this difference, however, is not significant, see Appendix.

In general, the magnitude of the relative dielectric constant $\varepsilon$ depends on which part of the medium relaxation is considered explicitly (as moving charges $q_i$) and which part is described phenomenologically as a polarizable dielectric[31]. Since in non-polarizable microscopic models the atomic motions are described explicitly, the screening factor should include only electronic component of the medium polarization, $\varepsilon = \varepsilon_{el}$. The static (i.e. time-independent) dielectric approximation in this case is quite accurate, because on the time scale of nuclear motion the electronic relaxation occurs almost instantaneously, reducing at once all interatomic electrostatic interactions by a factor of $\varepsilon_{el}$. The phenomenological parameter $\varepsilon_{el}$ can be theoretically evaluated by Clausius-Massotti equation (2.9) or directly measured as a high-frequency dielectric permittivity ($\varepsilon_{el} = n^2$, where $n$ is a refracting index of the medium). For organic materials typical values of $\varepsilon_{el}$ are in the range 1.7-2.2[10]. Thus, the uniform dielectric approximation with $\varepsilon_{el} = 2$ can be a good approximation for many biological systems. The resulting model, which combines a non-polarizable (fixed-charge) force field for nuclear dynamics (MD) and a phenomenological electronic continuum (EC) for the electronic polarization, is referred to as MDEC[29].

Summarizing, MDEC model considers charges $q_i$ moving in electronic polarizable continuum of known dielectric constant $\varepsilon_{el}$, see Fig.1a. In the uniform dielectric all electrostatic interactions are scaled by a factor $1/\varepsilon_{el}$. Since interactions are quadratic in charges, the effects of electronic dielectric screening can be taken into account implicitly by using scaled partial charges, $q_i^{eff} = q_i/\sqrt{\varepsilon_{el}}$. In the section "Applications of MDEC model" we show by the computational tests that the simple scaling procedure results in accurate effective interaction between ions of general shape in real solvents. The un-scaled original charges are difficult to specify a priory in general (they are not the same as partial atomic charges of a molecule in vacuum, see ref [29]), unless one deals with ions or ionized groups in proteins, whose un-scaled net charges are known. But effective charges $q_i^{eff}$ can be found empirically by fitting experimental data[32-33] or appropriately scaled ab initio interaction energies[2].

---

[‡] For example, for point dipoles in spherical cavities, the scaling factor for interaction is $[3\sqrt{\varepsilon}/(2\varepsilon+1)]^2$; for electronic dielectric constant of 2 (1.78 for water) the difference may not be significant, considering the uncertainty of the shape of molecular cavity. See details in Appendix.

**2.2 Solvation free energy**

In MDEC model, when the solvation free energy of a group is considered, the electronic polarization free energy is treated explicitly. The solvation free energy consists of the nuclear part $\Delta G_{nuc}$, evaluated by MD, and the pure electronic polarization part $\Delta G_{el}$ (which corresponds to the last term in Eq. (2.8)) evaluated by using the polarizable continuum model[34] (i.e. by solving the Poisson equation with corresponding boundary conditions, with dielectric constant $\varepsilon = 1$ inside the solute region and $\varepsilon = \varepsilon_{el}$ outside, as shown in Fig.1b):

$$\Delta G = \Delta G_{nuc} + \Delta G_{el}. \qquad (2.10)$$

The origin of the last term can be traced in the derivation of the previous sub-section, where the last term in Eq. (2.8) corresponds to $\Delta G_{el}$ for spherical ions. The derivation can be extended to the case of molecules of a general shape, in which case the electronic polarization energy for each molecule *a* takes a familiar reaction field energy form,

$$\Delta G_{el}^a = \frac{1}{2}\sum_i \varphi_{ai}^{RF} q_{ai}, \qquad (2.11)$$

where $\varphi_{ai}^{RF}$ are reaction field potentials at the position of molecular charges $q_{ai}$ for a given molecule, obtained by solving the corresponding Poisson equation, as mentioned above.

When the interaction of a solute with solvent molecules is considered in an MDEC simulation (in evaluating the $\Delta G_{nuc}$ part), the solute partial charges (found in an appropriate quantum-mechanical calculation, in vacuum or in a dielectric environment) should be scaled by $1/\sqrt{\varepsilon_{el}}$, like all other charges when the forces between atoms are considered. If no scaling of solute charges is employed in MD simulation, which is typical for standard MD technique, e.g. refs [7, 35], the free energies obtained from MD, $\Delta G_{MD}$, has to be corrected directly afterward. Since in the linear response approximation the solvation free energy is quadratic in charges of the solute, $\Delta G_{MD}$ should be corrected by a factor $1/\varepsilon_{el}$, giving $\Delta G_{nuc} = \Delta G_{MD}/\varepsilon_{el}$. The total MDEC polarization free energy of the medium then is:

$$\Delta G = \frac{1}{\varepsilon_{el}}\Delta G_{MD} + \Delta G_{el} \qquad (2.12)$$

where $\Delta G_{MD}$ is the electrostatic solvation free energy obtained in non-polarizable MD using un-scaled solute charges [§] (i.e. the standard approach), and $\Delta G_{el}$ is the pure electronic part of the free energy. A more detailed description of the free energy simulation technique accounting for the electronic polarization can be found in refs [27-29].

---

[§] The charges of the solvent molecules (such as water) are assumed to be already scaled, which is the case in standard AMBER[1], CHARMM[2], GROMOS[3] or OPLS[4] MD simulations; the empirical charges of *neutral* species, in contrast to charged species, are typically correctly reflect the condensed matter nature of the interaction.

Thus, unlike the common empirical models[1, 2-4], in MDEC model the electronic polarization term appears explicitly in the expression for the solvation free energy (2.12). The electronic component constitutes more than half of the solvation free energy for ions and affects the entire non-polarizable concept, including the parameterization strategy. The reason why the conventional non-polarizable force fields[1, 2-4] reproduce the hydration free energies of ions quite accurately, completely ignoring the electronic part of the solvation energy will be explained in the Section "Empirical force fields".

**2.3 Dielectric constant of the medium**

The dielectric constant of the medium is often employed in the continuum electrostatic e.g. for solvation free energy evaluation calculations[36]. In microscopic calculations, on the other hand, the solvation free energy is obtained directly from MD simulations. The question arises often as to what is the effective dielectric constant of the medium $\varepsilon_{MD}$ that corresponds to a specific microscopic model of the system. The free energy relationships discussed in the previous section allow one to make a connection between the total (static) dielectric constant, $\varepsilon_0$, which includes both nuclear and electronic polarization effects, and the dielectric constant of non-polarizable MD simulations, $\varepsilon_{MD}$, which does not explicitly describe pure electronic polarization of the medium.

Suppose we consider a spherical ion or a pair of spherical ions; in this case, according to Ref [37], the solvation energies will be proportional to their corresponding Born factors: $\Delta G \sim \left(1-\frac{1}{\varepsilon_0}\right)$, $\Delta G_{el} \sim \left(1-\frac{1}{\varepsilon_{el}}\right)$ and $\Delta G_{MD} \sim \left(1-\frac{1}{\varepsilon_{MD}}\right)$. From Eq.(2.12) we find:

$$\varepsilon_0 = \varepsilon_{MD} \cdot \varepsilon_{el} \qquad (2.13)$$

That is the total dielectric constant of the medium $\varepsilon_0$ is not equivalent to that reproduced by the (non-polarizable) MD simulation, $\varepsilon_{MD}$; instead, the relationship between the two is given by the above formula. Although the above arguments strictly valid only for spherical ions and for the bulk solvent modeled with periodic boundary conditions [38], the relation (2.13) is in fact more general.

The above relation can be also obtained using the well-known expression [38] for the static dielectric constant:

$$\varepsilon_0 = \varepsilon_{el} + \frac{4\pi}{3Vk_BT}\langle M^2 \rangle \qquad (2.14)$$

Here $\langle M^2 \rangle$ is the mean square fluctuation of the total dipole of the dielectric sample $V$; $k_B$ and $T$ are Boltzmann constant and temperature, respectively. According to the MDEC scaling procedure, the actual dipole moment $\mu$ of particles in the bulk is related to the effective moment $\mu^{eff}$ of these particles in non-polarizable model as $\mu = \sqrt{\varepsilon_{el}}\,\mu^{eff}$; therefore, $\langle M^2 \rangle = \varepsilon_{el}\langle M_{MD}^2 \rangle$; where $\langle M_{MD}^2 \rangle$ is the mean square fluctuation of the dipole moment observed in a non-polarizable MD. Thus, Eq.(2.13) is obtained from Eq.(2.14) by noticing that $\varepsilon_{MD}$ is defined via fluctuation $\langle M_{MD}^2 \rangle$ with $\varepsilon_{el}$=1 in Eq.(2.14).

## 3. Applications of MDEC model

### 3.1 Ab initio interactions modeled by charge scaling

According to MDEC model the electrostatic interactions between ions in the condensed phase should be reduced (scaled) by the factor $\varepsilon_{el}$, electronic dielectric constant, with respect to those in gas-phase. To test how well the simple charge scaling procedure reproduces the screening effect for spherical and non-spherical ions in electronic continuum with appropriate boundary conditions we considered[39] interaction of several charged species in ab initio calculations. The ab initio treatment captures the effects of electronic polarization of charged species themselves, while the effects of the polarization of the environment, and corresponding screening, are described here phenomenologically, by a continuum with dielectric $\varepsilon_{el} = 2.0$.

The interaction energies were calculated using a quantum-mechanical procedure identical to that of the CHARMM parameterization protocol [2]. For each structure with a given separation $r$ between moieties, the interaction energy was calculated as the difference between the total supermolecule energy and the sum of the individual monomer energies. The gas-phase interaction energies were calculated with model compounds in vacuum; while the bulk-phase interactions were obtained with the model compounds immersed in the dielectric of $\varepsilon = 2$. The quantum-mechanical calculation in dielectric utilized the PCM[34] technique and self-consistent reaction-field procedure implemented in Gaussian03 [40].

In Fig. 2 a, b, c, the ab initio interactions between ions $Na^+$–$Na^+$, $Arg^+$–$Glu^-$ and between $Glu^-$ ion and water are compared with those modeled by the original and scaled CHARMM [2] force fields. (For amino acids their corresponding model compounds are used.) In all cases, as expected, there is a significant screening effect of the dielectric environment on the interaction energy; the effect as seen, however, can be pretty accurately reproduced by a simple scaling of charges. Notice that the charges are scaled by a factor $\frac{1}{\sqrt{\varepsilon_{el}}}$; given what have been said about TIP3P water model (see Introduction, and the following sub-section on water), in the example of *Glu*⁻-$H_2O$ pair, only charges of *Glu*⁻ were scaled. Notice how accurately the scaled CHARMM force field reproduces the results of ab initio calculations. Similar results are expected for other common force fields (such as AMBER [1], GROMOS [3], etc) where charged groups are also treated as having their original vacuum net charges.

Thus, the simple charge scaling procedure in standard non-polarizable force fields can account for the effects of electronic screening not only in the interactions between ions but also between ions and water. Although, as seen in Fig.2 a, some additional adjustments of van-de-Waals parameters might be useful to improve the interactions at shorter distances.

### 3.2. Water models

Many non-polarizable force fields are essentially MDEC models. For example, TIP3P[32] or SPC/E[33] and similar models of water involve empirical charges that can be considered as scaled charges. TIP3P is particularly interesting in this regard as it is often used in biological simulations, and it serves as a reference for phenomenological parameters assignment of CHARMM [2].

It is known that the dipole moment of a water molecule in vacuum is 1.85D; in liquid state, however, the four hydrogen bonds to which each water molecule is exposed on average strongly polarize the molecule and its dipole moment becomes somewhere in the range[**] of 2.9D to 3.2D [42, 43]. The significant increase of the dipole from $\mu_0 = 1.85D$ to a value $\mu \approx 3D$, or even larger, is also supported by the Kirkwood-Onsager model [44], which estimates the enhanced polarization of a molecule due to the reaction field of the polarized environment. Yet, the dipole moment of TIP3P water model is only 2.35D. The specific value of TIP3P dipole moment can be understood as a scaled dipole, so that the dipole-dipole interactions are screened by the electronic continuum by a factor $1/\varepsilon_{el}$. Indeed, if each dipole (or all partial charges) is scaled by a factor $1/\sqrt{\varepsilon_{el}}$, one could consider interaction of the effective dipoles, $\mu^{eff} = \mu/\sqrt{\varepsilon_{el}} \simeq 2.35D$ (for water $\varepsilon_{el} = 1.78$), as if they were in vacuum. This appears to be exactly what the fixed-charge water models do. Thus, the charges of TIP3P water model should be understood as *scaled* charges that reflect the effect of electronic screening. (The scaling factor $1/\varepsilon_{el}$ for water dipoles is an approximation, and one can argue that a different scaling factor should be more appropriate, however, numerically all reasonable continuum models give about the same result, see Appendix.)

The scaled nature of charges of TIP3P water model becomes critically important when the interaction with a solute is considered. For example, if the charge of say $Na^+$ ion is assigned to be +1, then it is obviously inconsistent with the charges of water model; as the latter are scaled by a factor of $1/\sqrt{\varepsilon_{el}}$, while the charge of the ion is not. Clearly the strength of interaction is overestimated in this case by a missing factor of $1/\sqrt{\varepsilon_{el}}$, i.e. about 0.7 (for proteins $\varepsilon_{el} \sim 2$). The problem would not arise if the charge of the ion were appropriately scaled. (The reason why seemingly incorrect charge gives reasonable aqueous solvation free energy is explained in the next section.)

---

[**] It is recognized that in ab initio simulations of bulk water the water dipole can not be defined unambiguously and depends on the partitioning scheme used[41]; as such, its actual value remains a matter of debate. Here we rely upon calculations and the partitioning scheme of refs [42].

A detailed discussion of non-polarizable models such as TIP3P and SPC/E is given in our recent paper ref [45]. Here we briefly consider the transferability issue for water models that can be demonstrated using the Kirkwood-Onsager model.[44] In this model, the medium is represented by a continuum dielectric of $\varepsilon_0$ and the solvent molecule is modeled by a point polarizable dipole, placed in a spherical cavity of radius $R$; the permanent dipole is $\mu_0$ and the polarizability is $\alpha$. In such a model, for the average dipole moment of bulk water $\mu_l$ one obtains the following expression:

$$\mu_l = \frac{\mu_0}{1 - \frac{2(\varepsilon_0 - 1)}{(2\varepsilon_0 + 1)} \frac{\alpha}{R^3}}. \tag{3.1}$$

To use this model for estimation of the liquid state water dipole, one needs to know the value of the radius $R$ of the molecular cavity. This is obviously a phenomenological parameter, which needs to be fixed in comparison with experimental data or derived from a suitable theoretical model. One reasonable estimate of the radius of molecular sphere $R$ is to assume that $2R$ is the average distance between the liquid water molecules, $a=2R$. The distance between the molecules is related to their number density, $N_\alpha$, as $N_\alpha = a^{-3}=1/8R^3$. Employing now the Clausius-Massotti relation between $N_\alpha$ and $\varepsilon_{el}$ as given by Eq.(2.9) the radius $R$ is expressed as

$$R^3 = \frac{\pi}{6} \frac{(\varepsilon_{el} + 2)}{(\varepsilon_{el} - 1)} \alpha. \tag{3.2}$$

With the radius $R$ given by Eq.(3.2), the equilibrium dipole moment in the liquid phase becomes

$$\mu_l = \frac{\mu_0}{1 - \frac{6}{\pi} \frac{(\varepsilon_{el} - 1)}{(\varepsilon_{el} + 2)} \frac{2(\varepsilon_0 - 1)}{(2\varepsilon_0 + 1)}}, \tag{3.3}$$

which gives the value 3.0 D for the bulk water ($\varepsilon_{el}$ = 1.78, $\varepsilon_0$ = 78, $\mu_0$ =1.85D). This value is in good agreement with the recent experimental estimations; however, it is slightly different from the empirical SPC/E MDEC value, $2.35D\sqrt{1.78}$ =3.14D. This slight inconsistency can be corrected by the fine tuning of the model parameters in computational parameterization procedure[45].

The above model will now be used to examine the transferability of non-polarizable TIP3P or similar potentials for bio-molecular simulations in conditions which are quite different from that in pure liquid state of water. Although (3.3) is a crude model, it nevertheless qualitatively captures the electronic polarization effect induced by the polarizable environment. Taking $\mu_g$=1.855D for water dipole in vacuum[46], $\alpha$=1.47 Å$^3$ for polarizability[47] and the water cavity radius estimated in ref [45] as $R$ = 1.55 Å, one obtains an estimate of $\mu$ in different environments characterized by dielectric constant $\varepsilon_0$. The dependence of $\mu$ on $\varepsilon_0$ is shown in Fig.3.

In the high-dielectric region, $\varepsilon_0 \geq 20$, as seen, the water polarization is almost constant and similar to that of the water molecule in the bulk, $\varepsilon_0$ = 80. At smaller $\varepsilon_0$, however, the model indicates a significant dependence of the water dipole moment on the polarity of the environment. As shown in Fig.3, the dipole moment of a water molecule in the media with $\varepsilon_0$ <20 is significantly lower than the value $\mu_l$ of water in the bulk. Thus, in low-dielectric environments, such as proteins or membranes, water should be modeled using potentials different than those of TIP3P or SPC/E (the work to develop such models is in progress in our group at present).

### 3.3. Empirical force fields

In non-polarizable force fields of AMBER[1], CHARMM[2], GROMOS[3] or OPLS[4] the atomic partial charges of non-charged groups can be understood approximately as "scaled MDEC charges", because, as discussed above, these empirical parameters were chosen in such a way as to reflect the condensed matter nature (including screening) of the interaction. In contrast, the charges of ionized groups remain un-scaled and therefore do not reflect effects of electronic screening, and as such are treated as if they were in vacuum.

The non-polarizable TIP3P potential is particularly interesting, as it is often used in biological simulations, and it serves as a reference for phenomenological parameters assignment of CHARMM [2]. The scaled nature of charges of TIP3P water model is important to bear in mind when the interaction of such water models with a solute is considered. For example, if the net charge of say Glu⁻ ionized side chain is assigned to be −1 in simulation, then it is obviously inconsistent with the charges of water model, as the latter are scaled by a factor of $1/\sqrt{\varepsilon_{el}}$, while the charges of the ion are not. Clearly the strength of interaction is overestimated in this case by a missing factor $1/\sqrt{\varepsilon_{el}}$, i.e. about 0.7 (for proteins $\varepsilon_{el} \sim 2$). The problem would not arise if the charge of the ion were appropriately scaled.

In free energy simulations with non-polarizable force fields (and un-scaled charges), the pure electronic contribution to the electrostatic free energy is typically completely ignored, as e.g. in Refs. [7, 35]. Yet, in many cases such simulations pretty accurately reproduce experimental solvation energies; this may appear surprising, given the fact that about a half of the total solvation free energy (for charged solutes typically 50-100 kcal/mol) comes from electronic polarization of the medium. In fact, the neglect of large electronic polarization free energy is almost completely compensated by the use of "incorrect" bare solute charges in such simulations. This fortuitous compensation of errors, however, occurs only in the high-dielectric media, as can be seen from the following argument.

Consider for example Born solvation energy of $Na^+$ ion, $Q=+1$, in water; in standard simulations one would have approximately

$$\Delta G = \frac{Q^2}{2R}(1-\frac{1}{\varepsilon_{MD}}), \qquad (3.4)$$

where $\varepsilon_{MD}$ is the dielectric constant of water that corresponds to a specific MD model employed in the calculation. No matter which model of water is used, $\varepsilon_{MD}$ is much larger than unity, hence the overall estimate of the solvation free energy is $Q^2/2R$, which is independent of properties of the solvent, and can match pretty well the experimental value, provided the ionic radius $R$ is chosen correctly. The interaction between two charges consequently is taken to be then $Q^2/r$, completely disregarding the electronic screening of the interaction.

MDEC model suggests instead that in MD simulations the charge $Q$ should be scaled, *and* the electronic solvation free energy $\Delta G_{el}=\frac{Q^2}{2R}(1-\frac{1}{\varepsilon_{el}})$ added explicitly. In this case, the nuclear part of the free energy calculated in MD will be

$\frac{(Q/\sqrt{\varepsilon_{el}})^2}{2R}(1-\frac{1}{\varepsilon_{MD}})$ and the total free energy, Eq.(2.10), is given by

$$\Delta G = \frac{(Q/\sqrt{\varepsilon_{el}})^2}{2R}(1-\frac{1}{\varepsilon_{MD}})+\frac{Q^2}{2R}(1-\frac{1}{\varepsilon_{el}}) \qquad (3.5)$$

Since $\varepsilon_{MD} = \varepsilon_0 / \varepsilon_{el}$, as given by Eq.(2.13), the above expression correctly reproduces the expected result $(Q^2/2R)(1-1/\varepsilon_0)$. Notice that the charge is not scaled when the solvation is calculated in electronic continuum. Notice also, that it is only when $\varepsilon_{MD} \gg 1$, the two expressions (3.4) and (3.5) approximately give the same result. Yet, for interaction energy of two charges the MDEC gives the correct expression $Q^2/r\varepsilon_{el}$, while the standard approach gives $Q^2/r$.

It is seen that in high dielectric medium, the un-scaled charges result in twice as large contribution for the nuclear part of solvation energy compared with the correct value; as a result the missing electronic part, which is half of the total, exactly compensated.

Given that the un-scaled relation is only formally correct when

$$\varepsilon_{MD} = \varepsilon_0 / \varepsilon_{el} \gg 1, \qquad (3.6)$$

it is not surprising that the traditional non-polarizable approach works well in aqueous solutions ($\varepsilon_0/\varepsilon_{el}$ ~40), as e. g. in [7, 28]; however, the approach fails (i.e. significantly underestimates the polarization effects) in low dielectric media ($\varepsilon_0/\varepsilon_{el}$ ~1) as in refs [9-10, 13-14, 48-49].

### 3.4. Dielectric Properties

Here we consider the dielectric properties of polar (alcohols) and non-polar (alkanes) solvents published recently.[10, 21] The dielectric constants of these solvents were calculated by using both the conventional non-polarizable model and polarizable classical Drude oscillator model[10, 21]. A significant improvement was obtained when the effects of electronic polarization were included, in particular for non-polar alkanes. The results of such calculations are reproduced in Tables 1 and 2.

As far as the dielectric constant is concerned, MDEC model states that the static dielectric constant is simply a product of the value found in non-polarizable MD simulations and that of the electronic continuum, Eq.(2.13). As non-polarizable simulations have already been done, we use the published results and check the above relation. The results are shown in Table 1 and 2 for alcohols and alkanes, respectively, where MDEC model is compared with both polarizable Drude model and with experiment.[50]

To quantify the comparison, we introduce a parameter $\theta$, which is a measure of how well a given dielectric constant reproduces the results of charging free energy calculations of spherical ions. Since the free energies are proportional to the corresponding Born factors, $\Delta G \sim \left(1-\dfrac{1}{\varepsilon}\right)$, the parameter $\theta$ is defined as:

$$\theta = \frac{\left(1-1/\varepsilon^{sim}\right)-\left(1-1/\varepsilon^{exp}\right)}{\left(1-1/\varepsilon^{exp}\right)} \tag{3.7}$$

In Tables 1 and 2, parameter $\theta$ is shown for different types of simulations and for MDEC model.

As predicted by the criterion in Eq. (3.6) the traditional non-polarizable model satisfactorily reproduces the polarization effect in such polar media as alcohols (rmsd of $\theta < 4\%$, Table 1); although, the polarizable classical Drude oscillator model[21] and the MDEC model demonstrate much better agreement with the experiment (rmsd of $\theta < 1\%$ for both).

In the case of non-polar media the traditional MD approach completely fails (rmsd of $\theta \sim 100\%$, see Table 2); whereas, the polarizable Drude model[10] and the MDEC model satisfactorily describe the polarization of neat alkanes (rmsd of $\theta$ is 5.1% and 4.3%, respectively). The MDEC approach appears to be even slightly favorable in this case. Thus, in the above examples MDEC approach performs quite well for both polar and non-polar media.

We next explored the application of non-polarizable models to a low-dielectric interior of proteins, which have been studied in the past using standard MD simulations.[12] For the dielectric permeability of the most internal region of cytochrome *c*, and several other proteins, Simonson *et al.*[12, 15] reported a value around 1.5 (and even lower deeper inside). This value is apparently too low to be the actual dielectric constant of the protein; indeed it is lower than the pure electronic permeability $\varepsilon_{el} = 2$ estimated for cytochrome *c* in the polarizable calculation.[16] According to MDEC model, to obtain the total (static) dielectric constant, the results of non-polarizable simulations should be modified as given by Eq.(2.13). Considering the value 1.5 as corresponding to $\varepsilon_{MD}$, and using $\varepsilon_{el} = 2$, for static dielectric constant we obtain the value $\varepsilon_0 = 3.0$, which is in agreement with the value 2.9 estimated by Muegge *et al.*[51]

In a related work, the non-polarizable simulations have been used for the analysis of dielectric properties of the interior of redox protein Cytochrome *c* oxidase.[52] The charge insertion process has been studied that models deprotonation of His291 residue of $Cu_B$ catalytic center in (dehydrated) C*c*O. The free energy data and corresponding dielectric properties are given in Table 3. The reaction-field energy obtained in the traditional MD technique was found to correspond to an un-physically low protein dielectric constant of 1.3. However, when the electronic ($\varepsilon_{el} = 2.0$) polarization energy was added explicitly, as given by Eq.(2.12), the microscopic reaction-field energy could be reproduced with a more realistic value of protein dielectric of 2.6. The estimated magnitude of the dry C*c*O dielectric constant in the region of active site is consistent with earlier results for cytochrome *c* which are 2.9[51] or 3.0 (the value[15] corrected by Eq.(2.13)).

### 3.5. Solvation Free Energy

The comparison of the standard MD and MDEC simulations using GROMOS[3] force field with experimental data for the electrostatic hydration free energy of polyatomic ions is given in Fig.4. As seen, MDEC free energies reproduce experimental data within the experimental error which is typically about several kcal/mol for ions. As expected, the conventional non-polarizable MD also reasonably well reproduces the polarization effect in such a high dielectric media as liquid water.

The quality of free energy simulations, however, is different in the low-dielectric media such as liquid cyclohexane or protein interior of protein cytocrome *c* oxidase, which was described above, see Table 3. As predicted by the criterion in Eq. (3.6), the traditional non-polarizable MD completely fails in the simulations of low-dielectric environment. The obtained in ref [48] solvation free energies of Methyl and Propyl Guanidinium ions in cyclohexane are about zero (or of the order of thermal fluctuations ~$k_BT$), i.e. the standard non-polarizable MD simulations result in no polarization effect at all, $\varepsilon_{MD} \sim 1$, while free energies obtained[48] in polarizable MD are several orders of magnitude higher reflecting the correct dielectric constant of cyclohexane $\varepsilon_0 = 2.0$[50].

Thus, according to the criterion in Eq. (3.6) and simulation data summarized above the standard non-polarizable MD technique significantly underestimate the solvation free energy of ions and medium dielectric constant in the low-dielectric materials. In contrast, MDEC simulations are correct for both high and low-dielectric media.

**3.6. What Are the Microscopic Interactions In the Condensed Phase?**

To test how well the macroscopic electronic continuum approximation of MDEC model describes a molecular solvent, with its microscopic structure and corresponding inhomogeneity of electronic density, we examined[39] a model of two ions $A^-$ and $A^+$ dissolved in benzene. In the non-polar solvent the nuclear component of polarization is negligible and the total effect is almost exclusively determined by the electronic polarization therefore the total PMF accurately represents the electronic screening effect. The low-dielectric environment is similar to that in the interior of a protein, or a lipid membrane. The solvent now is described by the polarizable Drude oscillator model [22], whereas ions are treated by a standard non-polarizable force field (Coulomb and Lennard-Jones interactions; the LJ parameters for ions correspond to those of $Cl^-$ ion.)

For such a system, we calculate the electrostatic part of the potential of mean force and compare the results with those of scaled and un-scaled CHARMM calculation, using the concepts of MDEC theory, see Fig.5. The PMF gradient over $r$ gives the average electrostatic force acting between charged particles $A^-$ and $A^+$ in the bulk. The solvation free energy $\Delta G(r)$ of ions was evaluated by three alternative techniques: by polarizable MD, by the standard technique using non-polarizable CHARMM force field, and by using Eq.(2.10) and CHARMM force field with scaled ion charges, according to MDEC model.

As seen in Fig.5, when the space between ionic spheres is larger than a size of solvent molecules, the effects of solvent microscopic structure becomes unimportant, and the average interaction, both in polarizable and non-polarizable models of benzene, can be approximated by a simple Coulomb law with an effective dielectric constant (obviously the LJ interactions are not important in this region). In case of polarizable Drude oscillator model for solvent benzene, the average interaction between ions is reproduced with an effective dielectric constant $\varepsilon_0=1.88$.[††]

According to MDEC theory, Eq.(2.13), the total dielectric constant of the medium $\varepsilon_0$ is a product of the electronic dielectric $\varepsilon_{el}$ (due to Drude-polarization of benzene molecules) and that of nuclei, $\varepsilon_{MD}$. The latter was obtained in a separate simulation using non-polarizable CHARMM model of benzene; the corresponding value is $\varepsilon_{MD}=1.16$. According to Eq.(2.13) then, the corresponding electronic dielectric constant of polarizable model of benzene is $\varepsilon_{el}=\varepsilon_0/\varepsilon_{MD}=1.62$. As seen in Fig.5 (solid line), in perfect agreement with MDEC theory, the results of *polarizable* benzene simulations are reproduced by scaling charges of ions (by a factor $1/\sqrt{1.62}$) and running *non-polarizable* CHARMM simulations. Again, we see that all effects of electronic polarization can be incorporated by scaling charges of ions with a factor $1/\sqrt{\varepsilon_{el}}$.

---

[††] We notice that the experimental value of $\varepsilon_0$ for benzene is actually 2.3[50]; the underestimated value of $\varepsilon_0$ is a consequence of the reduced polarizability parameter employed in the benzene model[22], which is ~20% lower than experimental benzene polarizability. This lack of parameterization, however, is not essential for the model test provided by the consistent choice of the underestimated value of $\varepsilon_{el}$.

The significant deviation of the results of standard non-polarizable MD from those of polarizable and MDEC techniques shown in Fig.5, in fact, can be rationalized without the PMF simulations. Since scaling factor of each microscopic model is given by the corresponding dielectric constant (as defined in section 2.3), the PMF profiles are approximated by the corresponding Coulomb functions $\frac{-1}{\varepsilon_0 r}, \frac{-1}{\varepsilon_{MD} r}$ and $\frac{-1}{(\varepsilon_{el}\varepsilon_{MD})r}$ for the polarizable, non-polarizable CHARMM and MDEC techniques, respectively. Due to the relation (2.13), PMF functions for polarizable MD and MDEC should be the same, while, deviation from the CHARMM technique is estimated as $\left(1-\frac{1}{\varepsilon_{el}}\right)\frac{1}{\varepsilon_{MD} r}$. Thus, for the low-dielectric media where $\varepsilon_{el} = 2$ and $\varepsilon_{MD} \sim 1$, the deviation is $\sim\frac{1}{2r}$, which is significant even for larger separation distances (~16 kcal/mol for $r = 10$Å). In the high-dielectric media ($\varepsilon_{MD} \gg 1$), however, the difference will be much smaller. For instance, in water ($\varepsilon_{el} = 1.8$, $\varepsilon_{MD} \sim 100$ for TIP3P model[55]) the deviation will be just $\sim\frac{1}{225r}$, which is ~0.5 kcal/mol even for the shortest separation $r \leq 4$Å (contact ion pair: $r \leq 2R_{vdW}$). Since 0.5 kcal/mol is of the order of statistical uncertainty of MD the missing electronic screening effect is not noticeable in the standard non-polarizable simulations of water solutions.

Thus, despite a complex nature of electronic polarization in a real system, the effect can be described reasonably well in different solvation conditions by a simple charge scaling procedure; this opens a way to modify the standard force fields so as to improve the description of their charged groups by effectively incorporating the electronic screening of charges.

### 3.7. Dynamics of salt bridges in proteins

To demonstrate the significance of accounting for the electronic screening effect in protein dynamics simulations we modeled[39] fluctuations of an important salt bridge (*Arg*438−*PropD* of *heme a*$_3$, see the structure in ref [39]) in Cytochrome *c* Oxidase (C*c*O). This salt bridge (SB) controls water penetration to the hydrophobic (low-dielectric) cavity in the catalytic center of C*c*O [56-57]. The strength of the electrostatic interaction of the salt bridge determines the rate of its opening/closing and, as a result, the probability of water transfer to/from the catalytic cavity.

The distance *d* between O2D of Δ-*propionate* and 2HH2 of *Arg*438 has been chosen to characterize the fluctuations of the salt bridge gate during an MD run. The AMBER [1] force field was used. The distribution functions for distance *d* obtained with scaled and original un-scaled charges are shown in Fig.6a. Here no water in the cavity was included in the simulation.

It is seen that the SB dynamics becomes qualitatively different once electrostatic interactions between the charged *Arg*438$^+$ and the COO$^−$ group of Δ-*propionate* are reduced by a factor of $1/\varepsilon_{el}$ (in the simulations, $\varepsilon_{el}$=2.0). In contrast to the standard MD simulations [57], the fluctuations observed in the scaled model are significantly larger, so that the internal water can now easily pass through the opened SB gate, and enter the catalytic cavity. In fact, during an 5ns MD run with scaled charges, several such water transitions were observed.

In Fig.6b, the distribution functions of $d$ are shown from simulations that included water in (and around) the catalytic cavity of the enzyme. As we already pointed out, the electronic screening affects not only charge-charge interactions, but interaction with water as well. Here TIP3P model is taken without modification; the charge scaling affects only the salt-bridge groups. As seen in Fig.6b when the effects of electronic screening are included even more dramatic changes are observed.

Thus, standard (unscaled charges) MD simulations with and without water in the cavity lead to the conclusion that the salt bridge is formed 100% of the time; here stability of the salt bridge is quantified by the criterion $d < 3$Å, while, the bridge is observed only 98% or even 63% of the time in simulations with scaled charges without (see Fig.6a) and with water (see Fig.6b), respectively.

It is clear, that the account for electronic screening of charged groups can give rise to *qualitatively different* results in simulations of proteins. As we have shown[39], this can be achieved in a computationally effective way by simple charge scaling of ionized groups in the protein.

Unfortunately, there are no direct experimental data on the dynamics of the salt bridge discussed here to verify our proposal of electronic screening. However, as we argue in this paper such a scaling is obvious from theoretical point of view. An indirect comparison with an experiment, and support of charge scaling, is provided by some other computational studies, such as Zhu *at el*[58] where a heuristic approximation for the charge scaling of ionized side chains (variable dielectric constant ≥ 2) somewhat similar to ours was employed, which resulted in significant improvement in both side chain and loop prediction for protein conformations.

## 4. Conclusions

There is inconsistency in how the effects of electronic polarizability are treated in the commonly used non-polarizable empirical force-fields[1, 2-4]. The electronic screening effect, inherent for the condensed phase media, appears to be accounted for only for neutral moieties, whereas the charged residues are treated as if they were in vacuum. As a result, the electrostatic interactions between ionized groups are exaggerated by a factor of about 2. Also, an important electronic polarization term is typically neglected in simulations of solvation free energy or dielectric constant. The omission of the electronic contribution to the solvation energy is compensated by the exaggerated electrostatic interactions, but the complete compensation is possible only in high-dielectric media.

The discussed here non-polarizable MDEC (Molecular Dynamics in Electronic Continuum) model provides a theoretical framework for systematic accounting of the effects of electronic polarization, and suggests a modification of the standard non-polarizable force fields[1, 2-4] to make them consistent with the idea of uniform electronic screening of partial atomic charges. In a few examples, we compared the traditional non-polarizable MD simulations with MDEC simulations, and demonstrated how the charges of ionized groups can be rescaled to correspond to MDEC model. The present theory states that the charges of ionized groups of the protein, as well as charges of ions, in simulations with existing non-polarizable potentials such as CHARMM, AMBER, etc should be scaled; i.e. reduced by a factor $1/\sqrt{\varepsilon_{el}}$ (about 0.7), to reflect the electronic screening of the condensed medium relevant to biological applications. In the solvation free energy simulations the electronic part of the free energy (estimated by the continuum model) should be added explicitly to the nuclear part, which is obtained in non-polarizable MD simulations. The inclusion of electronic screening for charged moieties is shown to result in significant changes in protein dynamics and can give rise to new qualitative results compared with the traditional non-polarizable force fields simulations.

## Acknowledgments

It is our pleasure to acknowledge many insightful discussions with Dr Marshall Newton. This work has been supported in part by the NSF grant PHY 0646273, and NIH GM054052.

## Appendix

### Dielectric Screening

The reduction of electrostatic interactions in a dielectric medium with respect to that in vacuum is called the dielectric screening of interactions. The effective interaction between two localized groups of charges in a dielectric of $\varepsilon$ is defined via the potential of mean force (PMF):

$$U^{int}(r, \varepsilon) = \text{PMF} = G(r, \varepsilon) - G(r=\infty, \varepsilon), \qquad (A1)$$

here $r$ is an effective distance between two charge groups; $G(r, \varepsilon)$ is the total electrostatic free energy of the system; $G(r=\infty, \varepsilon)$ is the sum of free energies of individual groups. The dielectric screening factor $f(r, \varepsilon)$ then is defined as the ratio between interaction in dielectric and in vacuum:

$$f(r, \varepsilon) = U^{int}(r, \varepsilon) / U^{int}(r, \varepsilon=1). \tag{A2}$$

In general, a solution of the dielectric problem with appropriate boundary conditions is necessary to obtain the free energy $G(r, \varepsilon)$, effective interaction $U^{eff}(r, \varepsilon)$ and screening factor $f(r, \varepsilon)$. If molecules in the condensed medium are represented by the partial charges distributed inside the molecular cavity of general shape then the result obviously will depend not only on the dielectric property $\varepsilon$ and distance $r$ but also on the molecular charge distributions $\rho_1(\vec{x})$, $\rho_2(\vec{x})$ and cavity shapes $S_1(\vec{x})$, $S_2(\vec{x})$, where $\vec{x}$ stands for the Cartesian coordinates.

The complexity of the problem is significantly reduced in the case of spherical molecular cavities. First, consider two point dipoles $\vec{\mu}_1$ and $\vec{\mu}_2$ at centers of the spherical cuts in the dielectric of $\varepsilon$. The total free energy of the system is given by

$$G(r, \varepsilon) = \frac{1}{2} \int \Psi(\vec{x}; r, \varepsilon) \cdot \rho(\vec{x}) d^3x =$$

$$= \frac{1}{2} \int_{V_1} \Psi(\vec{x}; r, \varepsilon) \cdot \rho_1(\vec{x}) d^3x + \frac{1}{2} \int_{V_2} \Psi(\vec{x}; r, \varepsilon) \cdot \rho_2(\vec{x}) d^3x, \tag{A3}$$

where $\rho(\vec{x}) = \rho_1(\vec{x}) + \rho_2(\vec{x})$ is the total *external* charge distribution and $\Psi(\vec{x}; r, \varepsilon)$ is the total potential induced by the charge distribution. In the limit of large distances, $r \gg R$ - radius of cavities, the higher-order re-polarization of the cavities in response to each other polarization can be neglected and the solution for the 2-sphere problem is expressed via solutions for single sphere problems: $\Psi(\vec{x} \in V_1; r, \varepsilon) = \varphi_{in}^{(1)}(\vec{x}) + \varphi_{out}^{(2)}(\vec{x})$; $\Psi(\vec{x} \in V_2; r, \varepsilon) = \varphi_{out}^{(1)}(\vec{x}) + \varphi_{in}^{(2)}(\vec{x})$, where $\varphi_{in}^{(i)}$, $\varphi_{out}^{(i)}$ are the potentials induced by the charge distribution $\rho_i(\vec{x})$ inside and outside own cavity, respectively. Note that, strictly speaking, the potential $\varphi_{out}^{(i)}$ is not a solution for a single sphere problem. Thus, the free energy is partitioned onto:

$$G(r, \varepsilon) = G(r=\infty, \varepsilon) + \frac{1}{2}\int_{V_1} \varphi_{out}^{(2)}(\vec{x}) \cdot \rho_1(\vec{x}) d^3x + \frac{1}{2}\int_{V_2} \varphi_{out}^{(1)}(\vec{x}) \cdot \rho_2(\vec{x}) d^3x, \quad (A4)$$

where $G(r=\infty, \varepsilon)$ is the sum of free energies of individual groups:

$$G(r=\infty, \varepsilon) = \frac{1}{2}\int_{V_1} \varphi_{in}^{(1)}(\vec{x}) \cdot \rho_1(\vec{x}) d^3x + \frac{1}{2}\int_{V_2} \varphi_{in}^{(2)}(\vec{x}) \cdot \rho_2(\vec{x}) d^3x. \quad (A5)$$

We note that the last two terms in Eq.(A4) are equal due to symmetry of electrostatic interactions. Recalling now the expression for the charge density of dipole $\rho_i(\vec{x}) = -\vec{\mu}_i \cdot \vec{\nabla}\delta(\vec{x} - \vec{x}_i)$ and performing the integration in (A4) by parts we obtain the expression for $U^{int}(r, \varepsilon)$:

$$U^{int}(r, \varepsilon) = -\frac{1}{2}\left(\vec{\mu}_1 \cdot \vec{E}_{out}^{(2)}(\vec{x}_1) + \vec{\mu}_2 \cdot \vec{E}_{out}^{(1)}(\vec{x}_2)\right) \quad (A6)$$

here $\vec{E}_{out}^{(i)} = -\vec{\nabla}\varphi_{out}^{(i)}$ is the electric field induced by one dipole at the position of the other dipole. To obtain this field one need to remember that the electric field induced by the dipole outside own cavity $\vec{F}_{out}^{(i)}$ will be additionally modified inside the second cavity due to its re-polarization. At larger distances the dipole field $\vec{F}_{out}^{(i)}$ is approximately uniform and its modification inside the other cavity is given by the well known solution[59]: $\vec{E}_{out}^{(i)} = \frac{3\varepsilon}{2\varepsilon + 1}\vec{F}_{out}^{(i)}$. Recalling now the expression for the electric field of the point dipole created outside own cavity we obtain the resulting field at the position of the other dipole as:

$$\vec{E}_{out}^{(i)} = \frac{3\varepsilon}{2\varepsilon+1}\left[\frac{3}{2\varepsilon+1}\frac{3(\vec{\mu}_i\cdot\vec{r})\vec{r}-\vec{\mu}_i\cdot r^2}{r^5}\right] \quad (A7)$$

Thus, from Eq.(A6) we find the interaction between two ideal dipoles in spherical cavities:

$$U^{int}(r>>R,\,\varepsilon) = -\varepsilon\left(\frac{3}{2\varepsilon+1}\right)^2 \frac{3(\vec{\mu}_1\cdot\vec{r})(\vec{\mu}_2\cdot\vec{r})-(\vec{\mu}_1\cdot\vec{\mu}_2)\cdot r^2}{r^5} \quad (A8)$$

and the screening factor of electrostatic interactions defined by Eq.(A2) is

$$f(r>>R,\,\varepsilon) = \varepsilon\left(\frac{3}{2\varepsilon+1}\right)^2 \quad (A9)$$

The screening (A9), in fact, is true for arbitrary charge distributions $\rho_1(\vec{x})$, $\rho_2(\vec{x})$ in spherical cavities, for which the lowest multipole moment is the dipole. For a general distribution of charges $q_i$ the solution of the dielectric problem outside own cavity is given by the famous Kirkwood's expansion[60]:

$$\phi_{out}(\vec{r},\varepsilon) = \sum_{i=1}^{N} q_i \sum_{l=0}^{\infty} \frac{2l+1}{\varepsilon(l+1)+l} \frac{(x_i)^l}{r^{l+1}} P_l(\cos\theta_i) =$$

$$= \sum_{l=0}^{\infty} \frac{2l+1}{\varepsilon(l+1)+l}\left[\frac{4\pi}{2l+1}\frac{1}{r^{l+1}}\sum_{m=-l}^{l} q_{lm} Y_{lm}(\theta,\varphi)\right] \quad (A10)$$

here $\vec{x}_i$ are the positions of charges $q_i$ in respect to the sphere center, $\theta_i$ are the angles between $\vec{r}$ and $\vec{x}_i$ vectors, $P_l(z)$ and $Y_{lm}(\theta,\varphi)$ are the Legendre polynomials and spherical harmonics, respectively, and $q_{lm} = \sum_{i=1}^{N} q_i (x_i)^l Y_{lm}^*(\theta_i^s,\varphi_i^s)$ are spherical multipole moments[59] which are linearly related to the Cartesian multipole moments. In Eq.(A10) the vacuum component of the potential is grouped by brackets and the prefactor reflects the screening of the field corresponding to each multipole. It is seen that at larger distances $r \gg R$ the electric field is determined exclusively by moments $q_{Lm}$ of the lowest nonvanishing multipole $L$ and, therefore, the solution (A10) for the arbitrary charge distribution is equivalent to the field from the point multipole $q_{Lm}$ located at the center of the sphere. Thus, at larger distances $r \gg R$ the screening factor between arbitrary charge distributions in two spheres is the same as that between point charges ($L_1=L_2=0$), point dipoles ($L_1=L_2=1$) or any appropriate nonvanishing moments located at the center of these spheres.

The screening factors between charged ($L_1=0$) and dipolar ($L_2=1$) as well as between charged ($L_1=0$) and charged ($L_2=0$) distributions are obtained in similar manner giving the following combination rules for the dielectric screening:

$$f_{qq}(r \gg R, \varepsilon) = \frac{1}{\varepsilon} \qquad \text{(charge-charge)}$$

$$f_{qd}(r \gg R, \varepsilon) = \frac{3}{2\varepsilon+1} \qquad \text{(charge-dipole)} \qquad (A11)$$

$$f_{dd}(r \gg R, \varepsilon) = \varepsilon\left(\frac{3}{2\varepsilon+1}\right)^2 \qquad \text{(dipole-dipole), see though refs [24, 61]}.$$

One should remember that the obtained scaling factors for charge-dipole or dipole-dipole interactions are only for specific models of molecular cavities – spheres, and for large distances, and therefore should not be taken directly as more appropriate than a straightforward factor $1/\varepsilon$, in particular because the shape of molecular cavity is ill-defined. However, it does show that the true scaling in some models can be different from a simple $1/\varepsilon$.

**Table 1. Dielectric Constant of Bulk Alcohols Simulated by Different MD Models at $T = 298.15$ K.**

| alcohol | $\varepsilon_0$, exp [a] | $\varepsilon_{MD}$, npol MD [b] | $\varepsilon_0$, pol. MD [c] | $\varepsilon_{el}$, pol. MD [c] | $\varepsilon_0$, MDEC [d] |
|---|---|---|---|---|---|
| MeOH | 32.61 | 17.2 | 30.1 | 1.5 | 25.8 |
| EtOH | 24.85 | 18.8 | 21.4 | 1.6 | 30.08 |
| 2-PrOH | 19.26 | 13.7 | 17.6 | 1.7 | 23.29 |
| 2-BuOH | 15.94 | 7.8 | 15.8 | 1.7 | 13.26 |
| 1-PrOH | 20.52 | 15.2 | 19.5 | 1.6 | 24.32 |
| 1-BuOH | 17.33 | 10.8 | 21.2 | 1.7 | 18.36 |
| rmsd of $\theta$ [e], % | – | 3.7 | 0.6 | – | 0.9 |

[a] experimental values[50];
[b] conventional non-polarizable MD model;[21]
[c] polarizable classical Drude oscillator model;[21]
[d] MDEC model, Eq.(2.13), where $\varepsilon_{MD}$ and $\varepsilon_{el}$ are taken from [b] and [c], respectively;
[e] The relative error of Born factor, Eq.(3.7).

**Table 2. Dielectric Constant of Bulk Alkanes Simulated by Different MD Models.**

| Alkane | $T$, K | $\varepsilon_0$, exp [a] | $\varepsilon_{MD}$, npol MD [b] | $\varepsilon_0$, pol. MD [c] | $\varepsilon_{el}$, pol. MD [c] | $\varepsilon_0$, MDEC [d] |
|---|---|---|---|---|---|---|
| ethane | 184.55 | 1.7595 | 1.014 | 1.707 | 1.697 | 1.721 |
| propane | 231.08 | 1.7957 | 1.015 | 1.798 | 1.768 | 1.795 |
| butane | 272.65 | 1.8098 | 1.016 | 1.801 | 1.774 | 1.802 |
| isobutane | 261.43 | 1.8176 | 1.015 | 1.905 | 1.823 | 1.850 |
| heptane | 298.15 | 1.9113 | 1.018 | 2.021 | 1.977 | 2.013 |
| heptane | 312.15 | 1.8904 | 1.018 | 1.976 | 1.933 | 1.967 |
| decane | 298.15 | 1.9846 | 1.020 | 2.118 | 2.066 | 2.106 |
| decane | 312.15 | 1.9668 | 1.019 | 2.128 | 2.074 | 2.113 |
| rmsd of $\theta$ [e], % | – | – | 96.4 | 5.1 | – | 4.3 |

[a] experimental values[50];
[b] conventional non-polarizable MD model;[10]
[c] polarizable classical Drude oscillator model;[10]
[d] MDEC model, Eq.(2.13), where $\varepsilon_{MD}$ and $\varepsilon_{el}$ are taken from [b] and [c], respectively;
[e] The relative error of Born factor, Eq.(3.7).

**Table 3. Solvation free energy of ions in low-dielectric media obtained by different MD models**

| SYSTEM | Methyl Guanidinium in cyclohexane[48] | Propyl Guanidinium in cyclohexane[48] | His291 in dehydrated Cytocrome c Oxidase[52] |
|---|---|---|---|
| FF | CHARMM | CHARMM | AMBER |
| $\Delta G_{MD}$, kcal/mol | -0.81 | -0.75 | -15.7 |
| $\Delta G_{tot}$, kcal/mol | -27.5[a] | -28.6[a] | -45.6[b] |
| $\varepsilon_{MD}$ | 1.0[c] | 1.0[c] | 1.3[c] |
| $\varepsilon_{el}$ | 2.0 | 2.0 | 2.0 |
| $\varepsilon_0$ | 2.0[50] | 2.0[50] | 2.6[c] |

[a] Polarizable MD simulations.
[b] MDEC, Eq.(2.12).
[c] Dielectric constant is estimated in continuum electrostatic calculations of the solvation free energy adjusting $\varepsilon_{MD}$ to reproduce $\Delta G_{MD}$ or $\varepsilon_0$ to reproduce $\Delta G$.

**Figure Legends**

**Fig.1.** a) MDEC model for the electrostatic interactions between solute (large crossed circles) and solvent (small crossed circles) charges moving in the electronic continuum of dielectric constant $\varepsilon_{el}$; the same electronic dielectric constant $\varepsilon_{el}$ is assigned for both the solvent and solute regions; b) MDEC model for estimation of the pure electronic part of the electrostatic solvation free energy $\Delta G_{el}$.

**Fig. 2.** Interaction energies between a) $Na^+$–$Na^+$ ions; b) $Arg^+$ and $Glu^-$ amino-acids; c) $Glu^-$ amino-acid and water. Open circles stand for the energies obtained in gas-phase HF/6-31(d) calculation; filled squares are for the same interactions but calculated in dielectric of $\varepsilon = 2.0$. The dashed lines represent interaction energies obtained by the standard CHARMM force field, and using TIP3P water model in c). The solid lines represent the interaction energies obtained by the CHARMM force field with scaled charges ($\varepsilon_{el} = 2.0$), and TIP3P water model in c).

**Fig. 3.** Dependence of water dipole on the polarity of the environment, as given by the Kirkwood-Onsager model[44], Eq.(3.1). The parameters are $\mu_g$=1.855D, $\alpha$=1.47 Å$^3$ and $R$ = 1.55 Å.

**Fig. 4.** Charging free energy of polyatomic ions in aqua solution simulated in the Ref[28]. Opened symbols correspond to the traditional MD simulations, while filled symbols stand for MDEC technique, Eq.(2.10). The experimental values (dashed line) were obtained as a difference of total solvation energies for a given ion[53] and its hydrocarbon counterpart[54]. Partial charges and geometry of ions were found in self-consistent reaction-field

quantum-chemical procedure[40]. Circles correspond to AM1 level of theory, while squares and triangles to RHF/6-31G**. Van-der-waals radii of G43A force field[3] scaled by the factor $\kappa$=0.9 (filled circles and squared) and $\kappa$=0.8 (filled triangle) were used to build the molecular cavity in MDEC computation of electronic free energy component.

**Fig. 5.** PMF for an ion pair $A^+$ and $A^-$ in benzene. The squares, circles and triangles stand for the results obtained with polarizable MD, non-polarizable CHARMM and CHARMM with scaled charges of the ions, respectively. Continuous curves are the least square fitting of the simulation points by the Coulomb function $-1/\varepsilon r$ (with the Ewald correction, see Ref [39]). For polarizable simulations (solid line), the effective dielectric constant $\varepsilon_0$=1.88; for non-polarizable simulations (dashed line), the dielectric constant $\varepsilon_{MD}$=1.16. The triangles correspond to *non-polarizable* CHARMM simulations with *scaled* charges by a factor $1/\sqrt{(\varepsilon_0/\varepsilon_{MD})}$ according to MDEC model.

**Fig. 6.** Distribution functions of the distance $d$ between O2D ($\Delta$-*propionate* of *heme* $a_3$) and 2HH2 (Arg438) of bovine C$c$O salt-bridge: a) no water in the catalytic cavity; b) 4 water molecules are added to the cavity. Dashed lines represent distributions obtained in the standard MD, while solid lines stand for the distributions obtained in the MD with scaled charges of the ionized groups.

**Fig. 1**

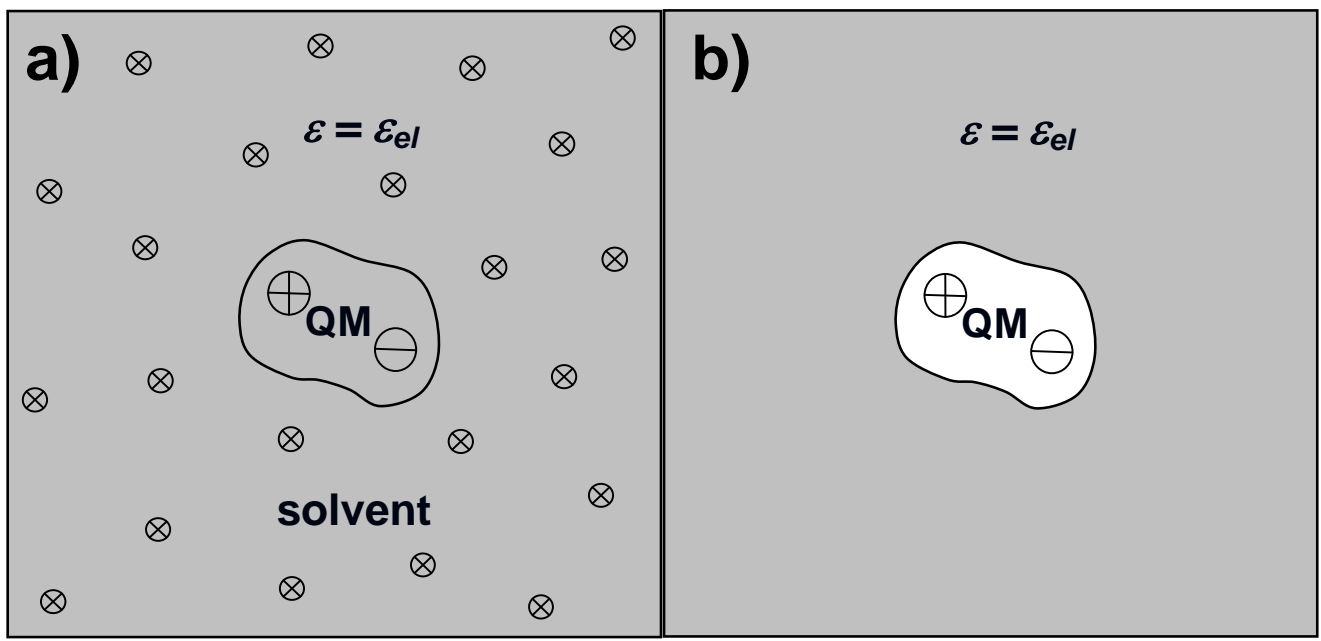

**Fig.2**

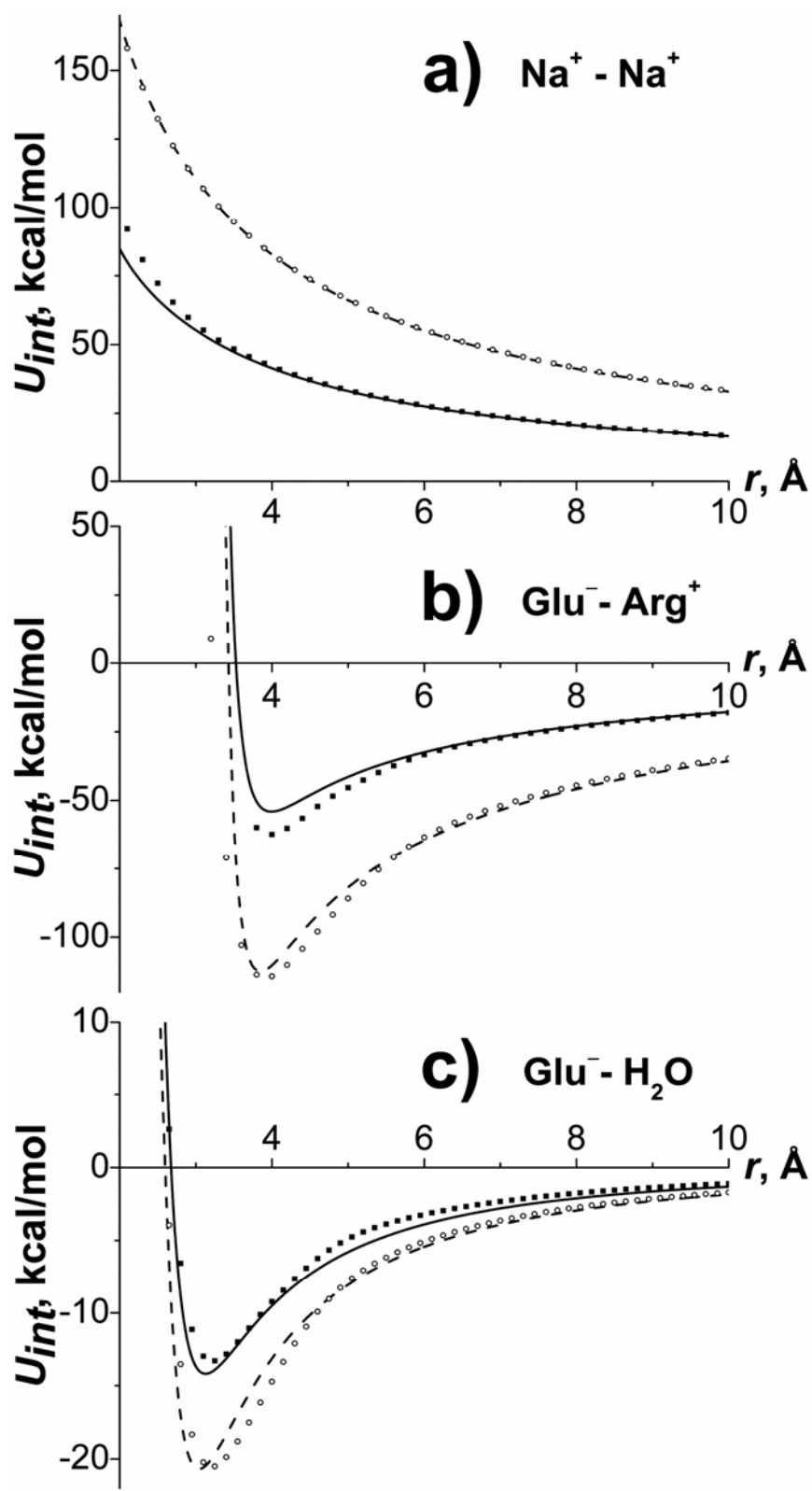

**Fig.3**

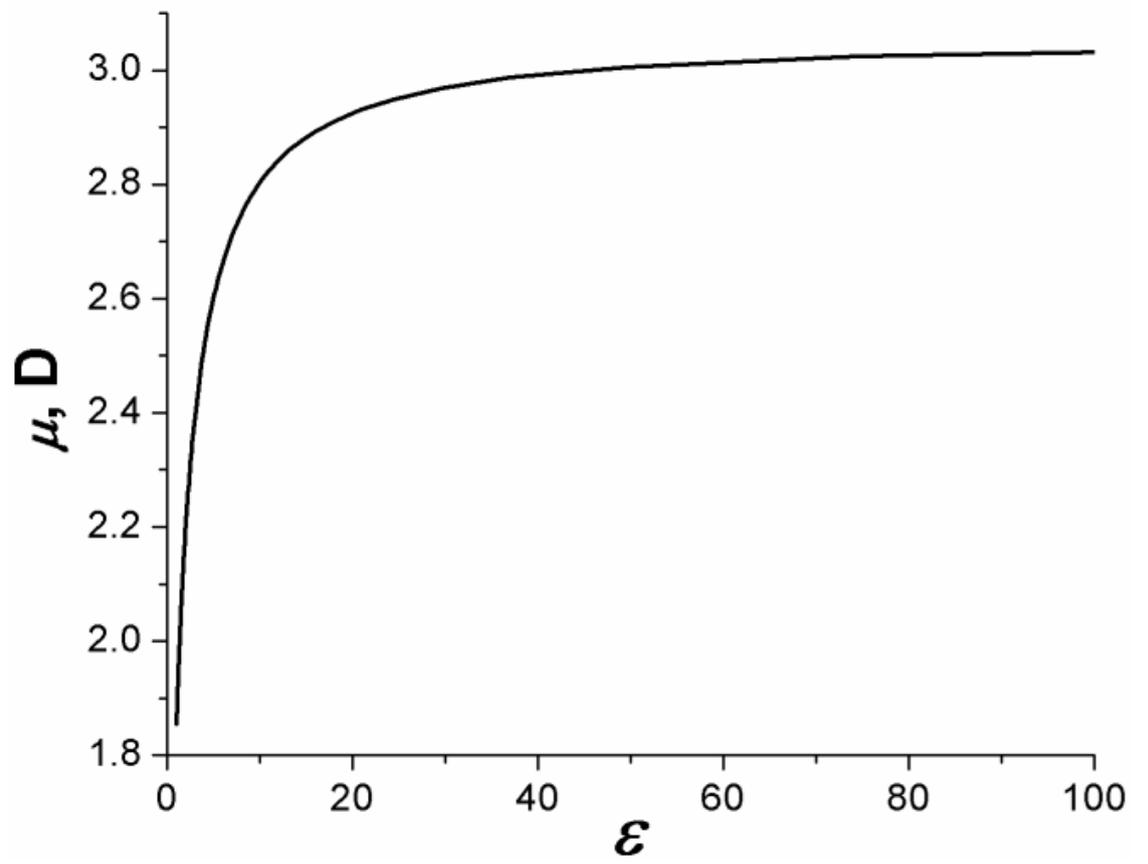

**Fig.4**

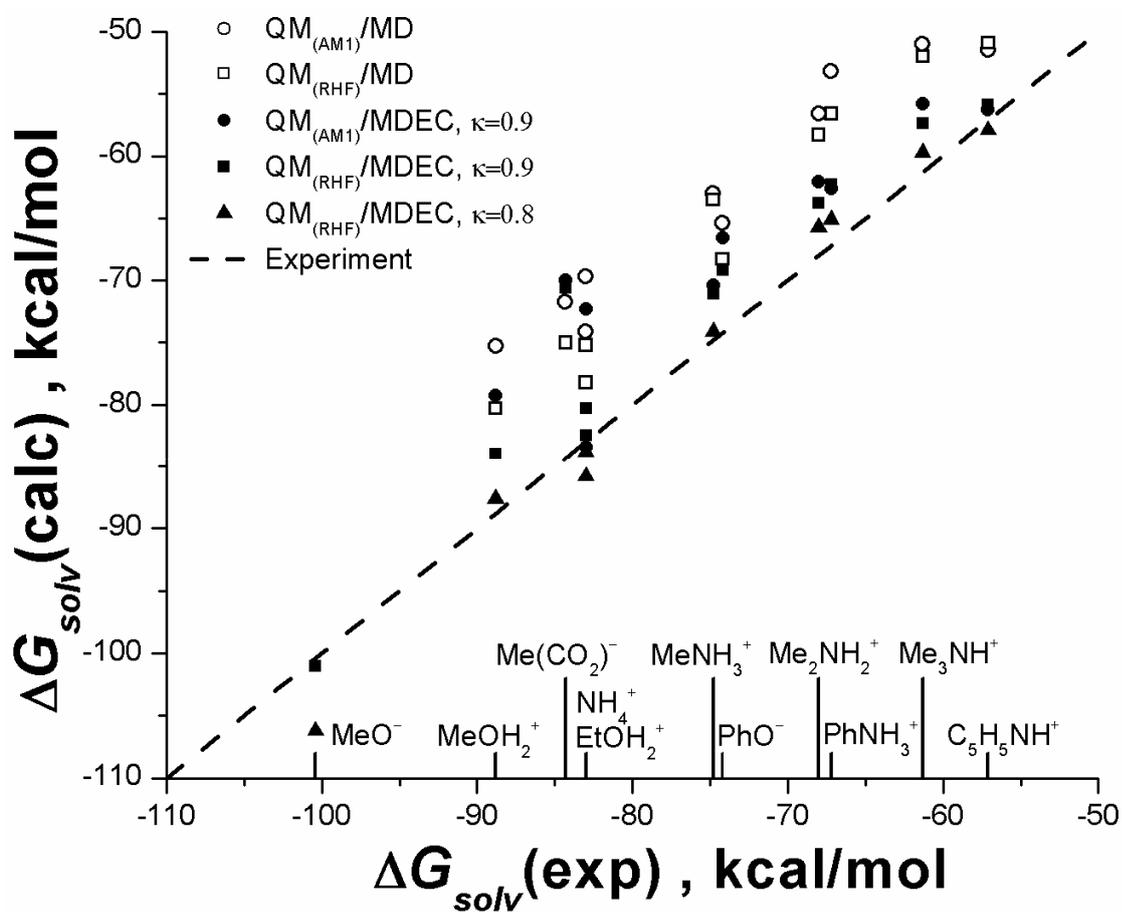

**Fig.5**

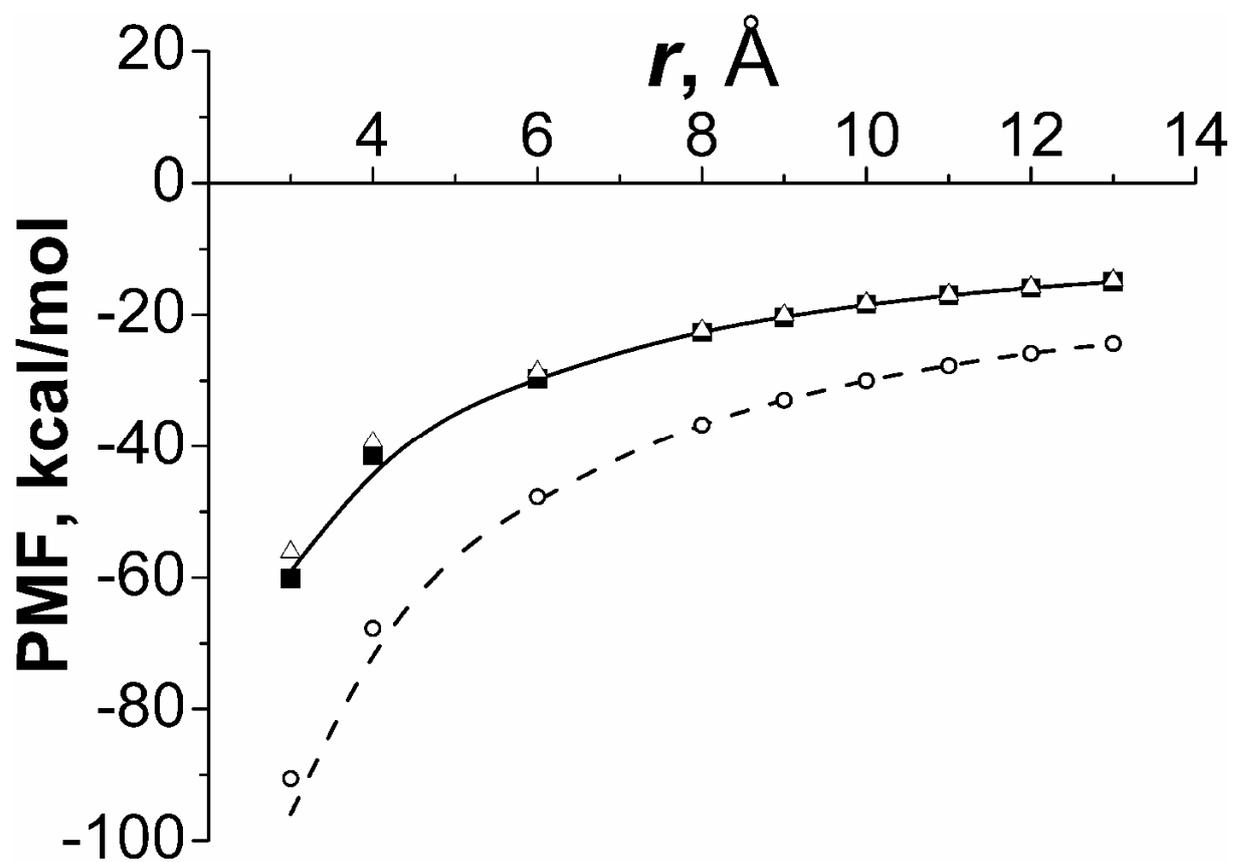

**Fig.6**

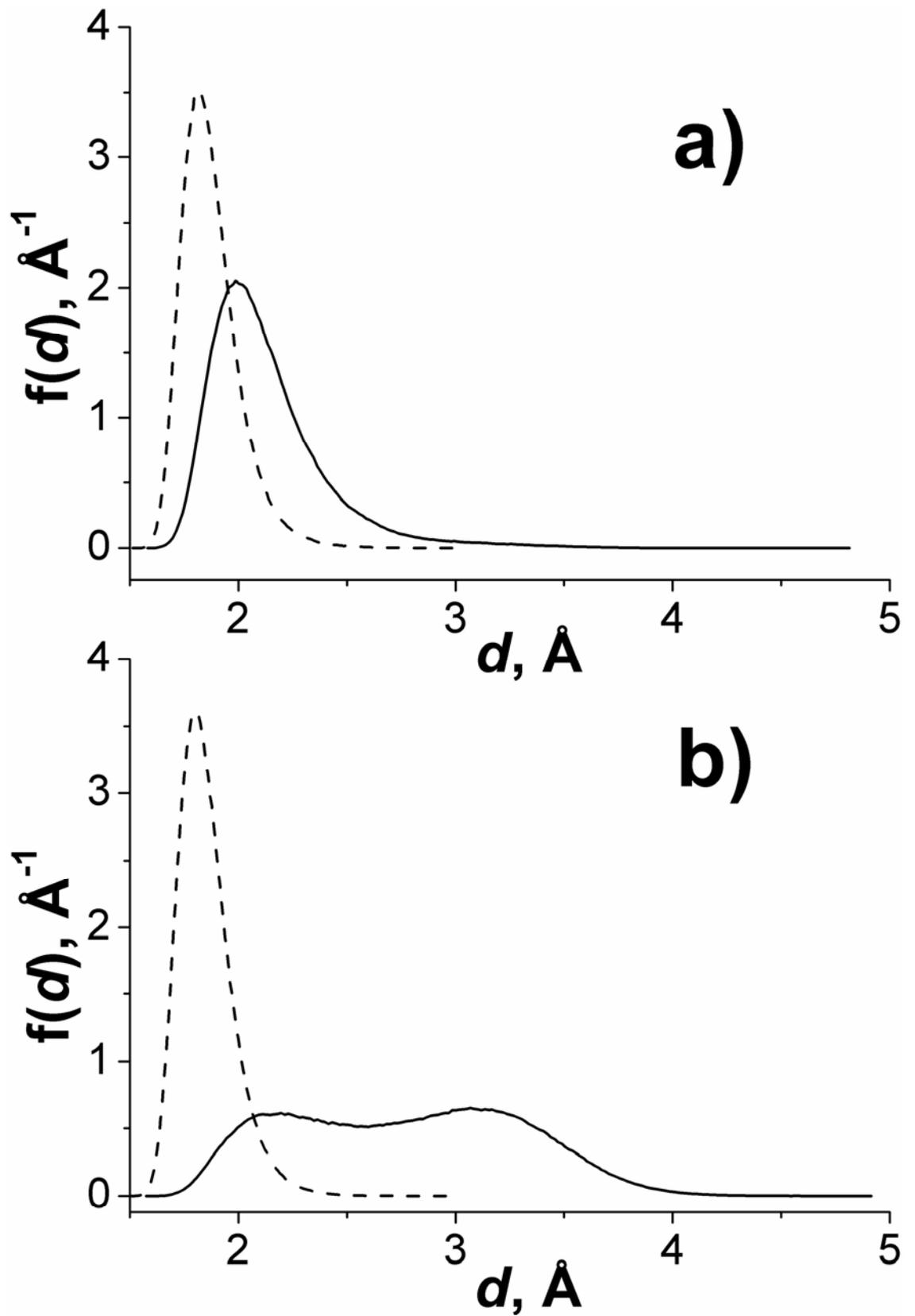